\documentclass{jnmp}

\usepackage{amsmath}
\usepackage{graphicx}
\JNMPnumberwithin{equation}{section}

\newtheorem{theorem}{Theorem}

\newtheorem{property}{Property}
\newtheorem{proposition}{Proposition}
\newtheorem{corollary}{Corollary}

\theoremstyle{definition}
\newtheorem{remark}{Remark}

\newtheorem*{example}{Example} % The '*' makes it unnumbered

\begin{document}

%  Headings
%
\renewcommand{\evenhead}{Y Tutiya}
\renewcommand{\oddhead}{Bright N--solitons for the INLS equation}

%  Titlepage
%
\thispagestyle{empty}

\FirstPageHead{*}{*}{20**}{\pageref{firstpage}--\pageref{lastpage}}{Article}
%  Parameters: Volume, number, year, page range, paper type
%  'Article' could be changed to 'Letter' or 'Review Article'

\copyrightnote{200*}{Y Tutiya}

\Name{Bright N--solitons for the intermediate nonlinear Schr\"{o}dinger equation}

\label{firstpage}

\Author{Yohei TUTIYA~$^\dag$}

\Address{$^\dag$ Graduate School of Mathematical Sciences, 
The University of Tokyo, 3-8-1 Komaba, Tokyo 153-8914, Japan\\
~~E-mail: tutiya@poisson.ms.u-tokyo.ac.jp}

\Date{Received Month *, 200*; Revised Month *, 200*; 
Accepted Month *, 200*}

\begin{abstract}
\noindent
  A previously unknown bright N-soliton solution for an intermediate nonlinear Schr\"{o}dinger equation of focusing type is presented.\ 
  This equation is constructed as a reduction of an integrable system related to a Sato equation of a 2-component KP hierarchy for certain differential--difference dispersion relations.\
Bright soliton solutions are obtained in the form of double Wronskian determinants.\
\end{abstract}

%  The paper
%
\section{Introduction}
\quad {This paper deals with the intermediate nonlinear Schr\"{o}dinger equation of focusing type and its bright soliton solutions.\
A general INLS equation can be written as follows.\
\begin{gather}
iu_t=u_{xx}+\sigma u(i+T)(|u|^2)_x\label{eq:originalINLS}
\end{gather}
Here, $u(x,t)$ is a complex function of two real variables $x,t$ and $T$ denotes the integral transformation
\begin{gather}
T[u](x)=\int_{-\infty}^\infty \hspace{-6.8mm}\backslash\hspace{3mm} \frac{1}{2\delta}\coth [\frac{\pi}{2\delta}(y-x)]u(y)dy.\label{eq:Ttransform}
\end{gather}
The integral sign with a backslash denotes the principal value.\ 
The constant $\sigma$ is taken to be $\pm 1$.\

For positive (negative) $\sigma$, the equation is called the defocusing (focusing) INLS equation.\
Originally, the defocusing INLS equation was discovered by carrying out a reductive perturbation method for the ILW equation~\cite{a}.\ 
It describes the long--term evolution of quasi--harmonic wave packets whose wavelength is short compared to the fluid depth.\ 

The inverse scattering transform for the defocusing equation and its Hirota bilinear form are known~\cite{g,h,i}.\ 
The dark soliton solutions of this equation have been also investigated intensively.\
But as for the focusing type, which is also an integrable system in its own right, as far as the author knows, soliton solutions have never come up in the literature.\ 
} 

In this paper, it will be shown that bright solitons exist for the focusing INLS equation and that they possess an interesting structure.\
An explicit formula for a bright $N$--soliton solution for this INLS equation will be presented.\
Our framework is basically a Sato description of the $2$-component KP hierarchy~\cite{b}, starting however from special differential--difference dispersion relations.\
Being related to the $2$--component KP hierarchy, the solutions of the resulting equations are automatically written as double Wronskians.\
This is one of the merits of this approach because it is generally a hard task to write $N$-soliton solutions in simple forms such as determinants.\   

{  Moreover, it is worth pointing out that the present situation runs parallel to the case of the NLS equation, for which it is known that the focusing NLS equation possesses bright soliton solutions of the double Wronskian type and, in contrast, the defocusing NLS equation possesses dark soliton solutions.\ 
%%%%%%%%%%%%%%%%%%%%%%%%%%%%%%%%%%%%%%%%%%
In fact, when the constant $\delta$ in (\ref{eq:originalINLS}) is regarded
as a deformation parameter for the equation, 
the NLS equation is recovered in the limit $\delta\to0$.\

%%%%%%%%%%%%%%%%%%%%%%%%%%%%%%%%%%%%%%%%%
{ 
In particular, for small $\delta$, the integral operator $T$ can be expanded as $T=-\frac1\delta \partial^{-1}+\frac\delta3\partial+o(\delta^3)$, and hence (\ref{eq:originalINLS}) can be written in the following form.\
\begin{gather}
iu_t=u_{xx}+\sigma u\left(-\frac1\delta +i\partial+\frac\delta3\partial^2+\cdots\right)|u|^2
\end{gather}
Thus, by setting $U=u/\sqrt{2\delta}$ and $\sigma=-1$, the focusing nonlinear Schr\"{o}dinger equation
\begin{gather}
iU_t=U_{xx}+2U|U|^2
\end{gather}
appears in the limit $\delta=0$.\
It will be shown later on that the bright soliton solutions for 
the INLS equation carry over to those of the 
NLS equation in this limit.\
This, in the author's point of view, justifies the use of the name ``INLS'' for the general equation (\ref{eq:originalINLS}).}
}

When discussing nonlocal soliton equations, one usually changes them first to differential--difference forms and treats them at that level.\ 
When one obtains a solution for the differential--difference system however, it is always problematic whether it is still a solution of the original nonlocal system.\
This problem often requires imposing certain analyticity conditions on the solutions, conditions which are highly nontrivial.\ 
As will be seen, the INLS equation is no exception.\

The structure of this paper is as follows.\ 
In section $2$, the INLS equation is decoupled into a certain differential--difference form.\ 
In section $3$, its linear problem is introduced.\ 
Next, it will be shown how to adapt Sato theory to differential--difference dispersion relations and a system of nonlinear evolution equations of Davey--Stewartson type is obtained.\
In section $4$, this system is reduced to the INLS equation in decoupled form and its $N$--soliton solutions are presented.\ 
In section $5$, it will be shown that these solutions are proper not only for the decoupled system but also the original nonlocal equation.\
%%%%%%%%%%%%%%%%%%%%%%%Decoupling the INLS equation%%%%%%%%%%%%%%%%%%%%%%%%
\section{Decoupling the INLS equation}
\quad The method that will be used to decouple the equation (\ref{eq:originalINLS}) into a differential--difference form owes largely to the properties of the integral transform (\ref{eq:Ttransform}) presented below~\cite{s,t}.\ 
As these are well-known facts, we shall skip the proofs.\   
\begin{property}\label{prop1}
Let $f(y)$ be a real function, define $F(z)$ as the integral
\begin{gather}
F(z):=\int_{-\infty}^\infty\frac{1}{2\delta}\coth [\frac{\pi}{2\delta}(y-z)]f(y)dy.
\end{gather} 

{$F(z)$ is a function defined in the entire complex $z$--plane, except for the horizontal lines ${\rm Im }z=2m\delta$, where $m$ runs over all the integers.\
Then, if $f(y)$ satisfies the H\"{o}lder condition on the real axis, the boundary values of $F(z)$ on $z=x$ ($x$ real) satisfy
\begin{gather}
F^\pm(x):=\lim_{\varepsilon\to 0^\pm}F(x+i\varepsilon)=T[f]\pm if(x). 
\end{gather}
In other word, $F^+(x)$ ($F^-(x)$) is the analytic continuation of $F(z)$ toward the real axis from the upper (lower) side}.\
In terms of the integral transformation (\ref{eq:Ttransform}), $F^+(x)$ and $F^-(x)$ also satisfy the relation\,  $F^+(x+2i\delta)=F^-(x)$.\ 
\end{property}
\begin{property}\label{prop2}
Suppose that a complex function $G(z)$ is analytic everywhere in the strip $0\leq \textrm{Im }z\leq 2\delta$ and is integrable on the real axis.\ 
(This condition will be referred to as ``{\it the analyticity condition}''  throughout the rest of the paper).\  
Then, the following equality holds for real $x$
\begin{gather}
iT[G(x+2i\delta)-G(x)]=G(x+2i\delta)+G(x).
\end{gather}
\end{property}

\vspace{\baselineskip}
Now, define $\ds v:=u^\dagger$ (where the dagger denotes complex conjugation) and 
\begin{gather}
w^\pm :=-\frac{1}{2}\lim_{\varepsilon\to 0^\pm}\int_{-\infty}^\infty\frac{1}{2\delta}\coth [\frac{\pi}{2\delta}(y-x\mp \varepsilon i)]|u(y)|^2dy,
\end{gather}
Then, (\ref{eq:originalINLS}) can be decoupled into the following system by means of {\rmfamily Property \ref{prop1}}.\
\begin{gather}
\left\{
\begin{array}{l}
\ds iu_t=u_{xx}-2uw^+_x,\\
\ds -iv_t=v_{xx}-2vw^-_x,\\
\ds w^--w^+=i\sigma uv.\
\end{array}
\right.\label{eq:ddform}
\end{gather}
Note that $w^+$ and $w^-$ satisfy $w^+(x+2i\delta)=w^-(x)$.\ 
In the next section, this system will be shown to be related to a $2$--component hierarchy.\ 
However it should be remembered that a solution for (\ref{eq:ddform}) will only be a solution to (\ref{eq:originalINLS}) if it satisfies {\it the analyticity condition}.\  
%%%%%%%%%%%%%%%%%%%%%%%Deispersion relation Sato equation%%%%%%%%%%%%%%%%%%
\section{Dispersion relations and Sato equations}
\quad Let us introduce the pseudo-differential operator,
\begin{gather}
P=I+W^{(1)}\partial^{-1}+W^{(2)}\partial^{-2}+\cdots+W^{(n)}\partial^{-n},\label{eq:PSD}
\end{gather}
where the coefficients $W^{(j)}\, (j = 1,2,\cdots )$ are $2\times 2$ matrices and $I$ is the unit matrix.\ 
In general, the $k,l$--elements of each $W^{(j)}$ are denoted as $w_{kl}^{(j)}$.\  
However, the elements of the first three matrices $W^{(1)},W^{(2)},W^{(3)}$ will be denoted as:
\begin{gather}
W^{(1)}=\left(\begin{array}{cc}w_{11}&w_{12}\\w_{21}&w_{22}\end{array}\right),\ W^{(2)}=\left(\begin{array}{cc}v_{11}&v_{12}\\v_{21}&v_{22}\end{array}\right),\
W^{(3)}=\left(\begin{array}{cc}u_{11}&u_{12}\\u_{21}&u_{22}\end{array}\right).
\end{gather}
Note that $w_{kl}^{(j)},w_{kl},v_{kl},u_{kl}$ are, apriori, different from $u,v,w$ in (\ref{eq:ddform}).\
$\partial^{j}$ denotes $( \partial /\partial\xi)^{j}$ for integer $j$.\
Though the theory is developed for the case of $n \to \infty$ in general, in this paper, we confine ourselves to (\ref{eq:PSD}) for simplicity~\cite{c,d} as the essence of the general theory is still kept in this simplification.

Let us consider the ordinary differential equation, 
\begin{gather}
P\partial^n\left(\begin{array}{c}f\\g\end{array}\right)=0.\label{eq:linearproblem}
\end{gather}
The $2n$ linearly independent solutions of (\ref{eq:linearproblem}) we denote $\ds \left(\begin{array}{c}f_1\\g_1\end{array}\right),\left(\begin{array}{c}f_2\\g_2\end{array}\right),\cdots,\left(\begin{array}{c}f_{2n}\\g_{2n}\end{array}\right)$.\ 
Now we assume that $W^{(j)},f_j,g_j$, in addition to $\xi$, depend on infinitely many continuous variables $t_1^{(1)},t_1^{(2)},t_2^{(1)},t_2^{(2)},t_3^{(1)},t_3^{(2)},\cdots$ and on two discrete variables $z_1,z_2$, by requiring $f_j$ and $g_j$ to satisfy the following dispersion relations
\begin{gather}
\ds \frac{\partial}{\partial t_k^{(1)}}\left(\begin{array}{c}f_j\\g_j\end{array}\right)=E_1\partial^k\left(\begin{array}{c}f_j\\g_j\end{array}\right),\quad
\ds\frac{\partial}{\partial t_k^{(2)}}\left(\begin{array}{c}f_j\\g_j\end{array}\right)=E_2\partial^k\left(\begin{array}{c}f_j\\g_j\end{array}\right)
,\label{eq:difdisp}\\[3mm]
\begin{array}{l}
\ds i\left\{\left(\begin{array}{cc}f_j^{\langle 1\rangle}\\g_j^{\langle 1\rangle}\end{array}\right)-\left(\begin{array}{cc}f_j\\g_j\end{array}\right)\right\}
=E_1\partial \left(\begin{array}{cc}f_j\\g_j\end{array}\right)
,\\[7mm]
i\left\{\left(\begin{array}{cc}f_j^{\langle 2\rangle}\\g_j^{\langle 2\rangle}\end{array}\right)-\left(\begin{array}{cc}f_j\\g_j\end{array}\right)\right\}
=E_2\partial \left(\begin{array}{cc}f_j\\g_j\end{array}\right),
\end{array}
\label{eq:shiftdisp}
\end{gather}
for ${}^\forall k,j$, where
\begin{gather}
E_1=\left(\begin{array}{cc}1&0\\0&0\end{array}\right),\ E_2=\left(\begin{array}{cc}0&0\\0&1\end{array}\right),
\end{gather}
and $\delta$ is a real and positive parameter.\ 
The symbol $\langle \mu\rangle (\mu=1,2)$ on top of a function means that $z_\mu$ is shifted forward by $2i\delta$.\ 
Note that the dispersion relation (\ref{eq:difdisp}) is the same as for $2$--component KP hierarchy~\cite{b}.\ 

Differentiating (\ref{eq:linearproblem}) with respect to $t_j^{(\mu)}$, we have the Sato equation 
\begin{gather}
\frac{\partial P}{\partial t_j^{(\mu)}}=B_j^{(\mu)}P-PE_\mu\partial^j.\label{eq:Sato1}
\end{gather}
The operator $B_j^{(\mu)}$ is defined by,
\begin{gather}
B_j^{(\mu)}:=\left(PE_\mu\partial^jP^{-1}\right)_+,
\end{gather}
where $(A)_+$ denotes the non--negative differential part of the pseudo differential operator $A$.\ 
The calculations that lead from (\ref{eq:Sato1}) to (\ref{eq:ilns1}) are of course the same as those for the $2$--component KP hierarchy.\
(More detail is available in~\cite{e}).\ 
The first few $B_j^{(\mu)}$'s are explicitly written as,
\begin{gather}
\begin{array}{l}
\ds B_1^{(1)}=E_1\partial + [W^{(1)},\ E_1],\quad 
B_1^{(2)}=E_2\partial + [W^{(1)},\ E_2]\\[3mm]
\ds B_2^{(1)}=E_1\partial^2 + [W^{(1)},\ E_1]\partial-W^{(1)}_{t_1^{(1)}}-E_1W^{(1)}_{\xi}\\[3mm]
\ds B_2^{(2)}=E_2\partial^2 + [W^{(1)},\ E_2]\partial-W^{(1)}_{t_1^{(2)}}-E_2W^{(1)}_{\xi}.
\end{array}
\end{gather}
%%%%%%%%%%%%%%%Lowest order equation%%%%%%%%%%%%%%%%%%%%%%%%%%%%%%%

The lowest order equations for $W^{(1)},W^{(2)},W^{(3)}$ are written below.\ Following Kac and van de Leur~\cite{e}, we write $W$ for $W^{(1)}$ to simplify the notation.\
\begin{gather}
\frac{\partial W}{\partial t_1^{(\mu)}}=E_\mu W_\xi+ [W,\ E_\mu]W-[W^{(2)},\ E_\mu]\label{eq:firstSato}\\
\frac{\partial W^{(2)}}{\partial t_1^{(\mu)}}=E_\mu W_\xi^{(2)}+ [W,\ E_\mu]W^{(2)}-[W^{(3)},\ E_\mu]\label{eq:discreteSato}\\
\begin{array}{l}
\ds\frac{\partial W}{\partial t_2^{(\mu)}}=-[W^{(3)},\ E_\mu]+E_\mu(2W^{(2)}_\xi+W_{\xi\xi})+[W,\ E_\mu](W^{(2)}+W_\xi)\\
\ds\hspace{6cm}-(W_{t_1^{(\mu)}}+E_\mu W_\xi)W.
\end{array}\label{eq:secondSato}
\end{gather}
In particular, some of the matrix elements of (\ref{eq:firstSato}) can be written explicitly as follows:
\begin{gather}
\begin{array}{ll}
\ds\frac{\partial w_{21}}{\partial t_1^{(1)}}=w_{21}w_{11}-v_{21},
&\ds\frac{\partial v_{21}}{\partial t_1^{(1)}}=w_{21}v_{11}-u_{21},\\
\ds\frac{\partial w_{12}}{\partial t_1^{(2)}}=w_{12}w_{22}-v_{12},
&\ds\frac{\partial v_{12}}{\partial t_1^{(2)}}=w_{12}v_{22}-u_{12},\\
\ds\frac{\partial w_{22}}{\partial t_1^{(1)}}=w_{21}w_{12},
&\ds\frac{\partial v_{22}}{\partial t_1^{(1)}}=w_{21}v_{12},\\
\ds\frac{\partial w_{11}}{\partial t_1^{(2)}}=w_{12}w_{21},
&\ds\frac{\partial v_{11}}{\partial t_1^{(2)}}=w_{12}v_{21}.
\end{array}\label{eq:firstSato1}
\end{gather}
Using (\ref{eq:firstSato1}), the (1,2) and (2,1)--components of (\ref{eq:secondSato}) can be written as   
\begin{gather}
\begin{array}{ll}
\ds\frac{\partial w_{21}}{\partial t_2^{(1)}}=-\frac{\partial^2w_{21}}{\partial t_1^{(1)^2}}+2w_{21}\frac{\partial w_{11}}{\partial t_1^{(1)}},
&\ds\frac{\partial w_{12}}{\partial t_2^{(1)}}=\frac{\partial^2 w_{12}}{\partial t_1^{(1)^2}}-2w_{12}\frac{\partial w_{11}}{\partial t_1^{(1)}},\\
\ds\frac{\partial w_{21}}{\partial t_2^{(2)}}=\frac{\partial^2w_{21}}{\partial t_1^{(2)^2}}-2w_{21}\frac{\partial w_{22}}{\partial t_1^{(2)}},
&\ds\frac{\partial w_{12}}{\partial t_2^{(2)}}=-\frac{\partial^2 w_{12}}{\partial t_1^{(2)^2}}+2w_{12}\frac{\partial w_{22}}{\partial t_1^{(2)}}.
\end{array}\label{eq:ilns1}
\end{gather}
Extending this construction, let us consider a difference operator with respect to $z_\mu$ acting on (\ref{eq:linearproblem}), which yields the Sato equation
\begin{gather}
i(P^{\langle \mu\rangle}-P)=C_\mu P-P^{\langle \mu\rangle}E_\mu\partial,\label{eq:Sato2}\\
C_\mu=\left(P^{\langle \mu\rangle}E_\mu\partial P^{-1}\right)_+=E_\mu\partial +W^{\langle\mu\rangle}E_\mu-E_\mu W.
\end{gather}
The coefficient of $\partial^{-1}$ in (\ref{eq:Sato2}) is 
\begin{gather}
i(W^{\langle\mu\rangle}-W)=E_\mu W^{(2)}-(W^{(2)})^{\langle\mu\rangle}E_\mu+E_\mu W_\xi+(W^{\langle\mu\rangle}E_\mu-E_\mu W)W\label{eq:descSato}
\end{gather}
And in particular, when $\mu=1$, the (2,2)--component of (\ref{eq:descSato}) is 
\begin{gather}
 i(w^{\langle 1\rangle}_{22}-w_{22})=w^{\langle 1\rangle}_{21}w_{12}.\label{eq:ilns5}
\end{gather}
and for $\mu=2$, the (1,1)--component of (\ref{eq:descSato}) is
\begin{gather}
i(w^{\langle 2\rangle}_{11}-w_{11})=w^{\langle 2\rangle}_{12}w_{21}.\label{eq:ilns6}
\end{gather}

Thus, from (\ref{eq:ilns1}),(\ref{eq:ilns5}) and (\ref{eq:ilns6}),  one obtains two systems of time evolution equations
\begin{gather}
\left\{
\begin{array}{l}
\ds\frac{\partial w_{21}}{\partial t_2^{(1)}}=-\frac{\partial^2w_{21}}{\partial t_1^{(1)^2}}+2w_{21}\frac{\partial w_{11}}{\partial t_1^{(1)}}\\[5mm]
\ds\frac{\partial w^{\langle2\rangle}_{12}}{\partial t_2^{(1)}}=\frac{\partial^2 w^{\langle2\rangle}_{12}}{\partial t_1^{(1)^2}}-2w^{\langle2\rangle}_{12}\frac{\partial w^{\langle2\rangle}_{11}}{\partial t_1^{(1)}}\\[5mm]
\ds i(w^{\langle2\rangle}_{11}-w_{11})=w^{\langle2\rangle}_{12}w_{21}
\end{array}
\right.\label{eq:mdavy1}
\end{gather}
and
\begin{gather}
\left\{
\begin{array}{l}
\ds\frac{\partial w_{12}}{\partial t_2^{(2)}}=-\frac{\partial^2 w_{12}}{\partial t_1^{(2)^2}}+2w_{12}\frac{\partial w_{22}}{\partial t_1^{(2)}}\\[5mm]
\ds\frac{\partial w^{\langle1\rangle}_{21}}{\partial t_2^{(2)}}=\frac{\partial^2w^{\langle1\rangle}_{21}}{\partial t_1^{(2)^2}}-2w^{\langle1\rangle}_{21}\frac{\partial w^{\langle1\rangle}_{22}}{\partial t_1^{(2)}}\\[5mm]
\ds i(w^{\langle1\rangle}_{22}-w_{22})=w^{\langle1\rangle}_{21}w_{12}
\end{array}
\right.\ .\label{eq:mdavy2}
\end{gather}
Because (\ref{eq:mdavy1}) and (\ref{eq:mdavy2}) are essentially the same system, we shall only deal with (\ref{eq:mdavy1}) hereafter.\
Note that (\ref{eq:mdavy1}) still has three independent variables though it looks very much like (\ref{eq:ddform}).\ 
%%%%%%%%%%%%%%%%%%%%%%%%%%The reduction to the INLS%%%%%%%%%%%%%%%%%%%%%
\section{The reduction to the INLS equation and its solutions}
\quad Introducing the change of variables $t_2^{(1)}=-it$, we impose the reduction condition,
\begin{gather}
\left(\frac{\partial}{\partial z_2}-\frac{\partial}{\partial t_1^{(1)}}\right)w_{ij}^{(k)}=0,\label{eq:red1}
%%%%the reduction condition for z_2=t_1^{(1)}%%%%
\end{gather}
which restricts the $t_1^{(1)},z_2$ dependence in $P$ to the form $t^{(1)}+z_2$.\
Hence, we introduce the new variable $x:=t^{(1)}+z_2$ with which (\ref{eq:mdavy1}) can be written as
\begin{gather}
\left\{
\begin{array}{l}
\ds iw_{21,t}=w_{21,xx}-2w_{21}w_{11,x}\\[1mm]
\ds -i\bar{w}_{12,t}=\bar{w}_{12,xx}-2\bar{w}_{12}\bar{w}_{11,x}\\[1mm]
\ds i(\bar{w}_{11}-w_{11})=\bar{w}_{12}w_{21}
\end{array}
\right.\label{eq:INLS1}
\end{gather}
where, $\bar{\cdot}$ denotes a $2i\delta$--shift with respect to $x$.\  
If, now, $w_{21}$ and $\bar{w}_{12}$ are complex conjugate for real $x$ and $t$ and if $w_{21}$ is analytic everywhere in the strip $0\leq \mbox{Im}\ x\leq 2\delta$,   
then (\ref{eq:INLS1}) can be represented as a single equation by means of {\rmfamily Property\ref{prop2}}
\begin{gather}
iw_{21,t}=w_{21,xx}-w_{21}(T+i)(|w_{21}|^2)_x.\label{eq:rINLS}
\end{gather}
Equation (\ref{eq:rINLS}) is the same as (\ref{eq:originalINLS}) for $\sigma=-1$.\ 
Concerning the last term of (\ref{eq:rINLS}), it also should be noticed that for the class of functions we are dealing with, the integral operator commutes with the $x$--derivative.\
Thus, the notations $(T+i)\{(|w_{21}|^2)_x\}$ and $\{(T+i)(|w_{21}|^2)\}_x$ represent the same function, which is denoted by $(T+i)(|w_{21}|^2)_x$.\    

%%%%%%%%%%%%%%%%%%%%%%%%Double Wronskian%%%%%%%%%%%%%%%%%%%%%%%%%%%%%%
Now we go on to consider actual solutions.\ 
Expressing equation (\ref{eq:linearproblem}) on the solutions $\left(\begin{array}{c}f_j\\g_j\end{array}\right) (j=1,2,\cdots,2n)$, one can write 
\begin{gather}
\left(\begin{array}{ccc|ccc}
f_1&\cdots&\partial^{n-1}f_1&g_1&\cdots&\partial^{n-1}g_1\\
f_2&\cdots&\partial^{n-1}f_2&g_2&\cdots&\partial^{n-1}g_2\\
\vdots&\vdots&\vdots&\vdots&\vdots&\vdots\\
f_{2n}&\cdots&\partial^{n-1}f_{2n}&g_{2n}&\cdots&\partial^{n-1}g_{2n}
\end{array}\right)
\left(\begin{array}{cc}
w_{11}^{(n)}&w_{21}^{(n)}\\
\vdots&\vdots\\
w_{11}^{(1)}&w_{21}^{(1)}\\\hline
w_{12}^{(n)}&w_{22}^{(n)}\\
\vdots&\vdots\\
w_{12}^{(1)}&w_{22}^{(1)}
\end{array}\right)\nonumber\\
\quad =
-\left(\begin{array}{cc}
\partial^nf_1&\partial^ng_1\\
\vdots&\vdots\\
\partial^nf_{2n}&\partial^ng_{2n}
\end{array}\right)
\end{gather}
Then, by means of Cramer's rule, $w_{21},w_{12},w_{11}$ are expressible as (in other words, a solution of (\ref{eq:mdavy1}) can be expressed as)
\begin{subequations}\label{wronskianw}
\begin{gather}
w_{21}=(-)^{n+1}\frac{\left|
\begin{array}{cccccccc}
f_1&\partial_\xi f_1&\cdots&\partial_\xi^{n-2} f_1&g_1&\partial_\xi g_1&\cdots&\partial_\xi^{n}g_2\\
f_2&\partial_\xi f_2&\cdots&\partial_\xi^{n-2}f_2&g_2&\partial_\xi g_2&\cdots&\partial_\xi^{n}g_2\\
\vdots&\vdots&\vdots&\vdots&\vdots&\vdots&\vdots&\vdots\\
f_{2n}&\partial_\xi f_{2n}&\cdots&\partial_\xi^{n-2}f_{2n}&g_{2n}&\partial_\xi g_{2n}&\cdots&\partial_\xi^{n}g_{2n}
\end{array}
\right|}{\left|
\begin{array}{cccccccc}
f_1&\partial_\xi f_1&\cdots&\partial_\xi^{n-1}f_1&g_1&\partial_\xi g_1&\cdots&\partial_\xi^{n-1}g_1\\
f_2&\partial_\xi f_2&\cdots&\partial_\xi^{n-1}f_2&g_2&\partial_\xi g_2&\cdots&\partial_\xi^{n-1}g_2\\
\vdots&\vdots&\vdots&\vdots&\vdots&\vdots&\vdots&\vdots\\
f_{2n}&\partial_\xi f_{2n}&\cdots&\partial_\xi^{n-1}f_{2n}&g_{2n}&\partial_\xi g_{2n}&\cdots&\partial_\xi^{n-1}g_{2n}
\end{array}
\right|}\label{eq:wronskianw21}
\end{gather}
\begin{gather}
w_{12}=(-)^{n}\frac{\left|
\begin{array}{cccccccc}
f_1&\partial_\xi f_1&\cdots&\partial_\xi^{n} f_1&g_1&\partial_\xi g_1&\cdots&\partial_\xi^{n-2}g_2\\
f_2&\partial_\xi f_2&\cdots&\partial_\xi^{n} f_2&g_2&\partial_\xi g_2&\cdots&\partial_\xi^{n-2}g_2\\
\vdots&\vdots&\vdots&\vdots&\vdots&\vdots&\vdots&\vdots\\
f_{2n}&\partial_\xi f_{2n}&\cdots&\partial_\xi^{n} f_{2n}&g_{2n}&\partial_\xi g_{2n}&\cdots&\partial_\xi^{n-2}g_{2n}
\end{array}
\right|}{\left|
\begin{array}{cccccccc}
f_1&\partial_\xi f_1&\cdots&\partial_\xi^{n-1}f_1&g_1&\partial_\xi g_1&\cdots&\partial_\xi^{n-1}g_1\\
f_2&\partial_\xi f_2&\cdots&\partial_\xi^{n-1}f_2&g_2&\partial_\xi g_2&\cdots&\partial_\xi^{n-1}g_2\\
\vdots&\vdots&\vdots&\vdots&\vdots&\vdots&\vdots&\vdots\\
f_{2n}&\partial_\xi f_{2n}&\cdots&\partial_\xi^{n-1}f_{2n}&g_{2n}&\partial_\xi g_{2n}&\cdots&\partial_\xi^{n-1}g_{2n}
\end{array}
\right|}\label{eq:wronskianw12}
\end{gather}
\begin{gather}
w_{11}=-\frac{\left|
\begin{array}{cccccccc}
f_1&\cdots&\partial_\xi^{n-2} f_1&\partial_\xi^{n} f_1&g_1&\partial_\xi g_1&\cdots&\partial_\xi^{n-1}g_2\\
f_2&\cdots&\partial_\xi^{n-2}f_2&\partial_\xi^{n} f_2&g_2&\partial_\xi g_2&\cdots&\partial_\xi^{n-1}g_2\\
\vdots&\vdots&\vdots&\vdots&\vdots&\vdots&\vdots\\
f_{2n}&\cdots&\partial_\xi^{n-2}f_{2n}&\partial_\xi^{n} f_{2n}&g_{2n}&\partial_\xi g_{2n}&\cdots&\partial_\xi^{n-1}g_{2n}
\end{array}
\right|}{\left|
\begin{array}{cccccccc}
f_1&\partial_\xi f_1&\cdots&\partial_\xi^{n-1}f_1&g_1&\partial_\xi g_1&\cdots&\partial_\xi^{n-1}g_1\\
f_2&\partial_\xi f_2&\cdots&\partial_\xi^{n-1}f_2&g_2&\partial_\xi g_2&\cdots&\partial_\xi^{n-1}g_2\\
\vdots&\vdots&\vdots&\vdots&\vdots&\vdots&\vdots&\vdots\\
f_{2n}&\partial_\xi f_{2n}&\cdots&\partial_\xi^{n-1}f_{2n}&g_{2n}&\partial_\xi g_{2n}&\cdots&\partial_\xi^{n-1}g_{2n}
\end{array}
\right|}\label{eq:wronskianw11}
\end{gather}
\end{subequations}
To describe the $n$--soliton solution we set $f_j,g_j$ 
\begin{gather}
\left\{
\begin{array}{l}
\ds f_j=c_j{\rm exp}\left[k_j\xi+k_jt_1^{(1)}+k_j^2t_2^{(1)}+a_j\right]\\[2mm]
\ds g_j={\rm exp}\left[-k_jz_2+i\left(e^{-2i\delta k_j}-1\right)\xi\right]
\end{array}\right.,\label{eq:solitonfg}
\end{gather}
where we have omitted independent variables which do not appear in (\ref{eq:INLS1}).\
{ 
We choose not to absorb the constant $c_j$
in the phase $a_j$ as it will take a specific value
in relation to {\it the analiticity condition} 
(cf. Proposition 4.1), whereas $a_j$ will essentially remain as a free parameter.\

}
Now, we divide the $j$th row of both the numerator and denominator of (\ref{eq:wronskianw21}) by $g_j$ (and similarly by $f_j$ for (\ref{eq:wronskianw12})).\ 
Then, we see that $w_{21}$, $w_{12}$ and $w_{11}$ contain $f_j$ and $g_j$ only in the form $f_j/g_j$ (or its inverse).\ 
Obviously $f_j/g_j$ contains $t_1^{(1)}$ and $z_2$ only in the form $x=t_1^{(1)}+z_2$.\ 
Thus, by setting 
\begin{gather}
\begin{array}{rl}
f_j/g_j=c_je^{\lambda_j}&:=c_j{\rm exp}\left\{\left(k_j-L_j\right)\xi+k_jx-ik_j^2 t+a_j\right\},\\[2mm]
L_j&:=i\left(e^{-2i\delta k_j}-1\right),
\end{array}
\end{gather}
$w_{21}$ and $w_{12}$ are expressible as follows.\ 
\begin{subequations}\label{eq:ourw}
\begin{gather}
w_{21}(x)=(-)^{n+1}\nonumber\\
\ \times\frac{ 
\left|
\begin{array}{cccccccc}
c_1e^{\lambda_1}&k_1 c_1e^{\lambda_1} &\cdots&k_1^{n-2} c_1e^{\lambda_1} &1&L_1&\cdots&L_1^{n}\\
c_2e^{\lambda_2}&k_2 c_2e^{\lambda_2} &\cdots&k_2^{n-2} c_2e^{\lambda_2} &1&L_2&\cdots&L_2^{n}\\
\vdots&\vdots&\vdots&\vdots&\vdots&\vdots&\vdots&\vdots\\
c_{2n}e^{\lambda_{2n}}&k_{2n} c_{2n}e^{\lambda_{2n}} &\cdots&k_{2n}^{n-2} c_{2n}e^{\lambda_{2n}} &1&L_{2n}&\cdots&L_{2n}^{n}\\
\end{array}
\right|}{
\left|
\begin{array}{cccccccc}
c_1e^{\lambda_1}&k_1 c_1e^{\lambda_1} &\cdots&k_1^{n-1} c_1e^{\lambda_1} &1&L_1&\cdots&L_1^{n-1}\\
c_2e^{\lambda_2}&k_2 c_2e^{\lambda_2} &\cdots&k_2^{n-1} c_2e^{\lambda_2} &1&L_2&\cdots&L_2^{n-1}\\
\vdots&\vdots&\vdots&\vdots&\vdots&\vdots&\vdots&\vdots\\
c_{2n}e^{\lambda_{2n}}&k_{2n} c_{2n}e^{\lambda_{2n}} &\cdots&k_{2n}^{n-1} c_{2n}e^{\lambda_{2n}} &1&L_{2n}&\cdots&L_{2n}^{n-1}\\
\end{array}
\right|}\label{eq:ourw21}
\end{gather}
\begin{gather}
w_{12}(x)=(-)^{n}\nonumber\\
\ \times\frac{\left|
\begin{array}{cccccccc}
1&k_1 &\cdots&k_1^{n} &c_1^{-1}e^{-\lambda_1}&L_1 c_1^{-1}e^{-\lambda_1}&\cdots&L_1^{n-2} c_1^{-1}e^{-\lambda_1}\\
1&k_2 &\cdots&k_2^{n} &c_2^{-1}e^{-\lambda_2}&L_2 c_2^{-1}e^{-\lambda_2}&\cdots&L_2^{n-2} c_2^{-1}e^{-\lambda_2}\\
\vdots&\vdots&\vdots&\vdots&\vdots&\vdots&\vdots&\vdots\\
1&k_{2n} &\cdots&k_{2n}^{n} &c_{2n}^{-1}e^{-\lambda_{2n}}&L_{2n} c_{2n}^{-1}e^{-\lambda_{2n}}&\cdots&L_{2n}^{n-2} c_{2n}^{-1}e^{-\lambda_{2n}}\\
\end{array}
\right|}{\left|
\begin{array}{cccccccc}
1&k_1 &\cdots&k_1^{n-1} &c_1^{-1}e^{-\lambda_1}&L_1 c_1^{-1}e^{-\lambda_1}&\cdots&L_1^{n-1} c_1^{-1}e^{-\lambda_1}\\
1&k_2 &\cdots&k_2^{n-1} &c_2^{-1}e^{-\lambda_2}&L_2 c_2^{-1}e^{-\lambda_2}&\cdots&L_2^{n-1} c_2^{-1}e^{-\lambda_2}\\
\vdots&\vdots&\vdots&\vdots&\vdots&\vdots&\vdots&\vdots\\
1&k_{2n} &\cdots&k_{2n}^{n-1} &c_{2n}^{-1}e^{-\lambda_{2n}}&L_{2n} c_{2n}^{-1}e^{-\lambda_{2n}}&\cdots&L_{2n}^{n-1} c_{2n}^{-1}e^{-\lambda_{2n}}\\
\end{array}
\right|}\label{eq:w12}
\end{gather}
\begin{gather}
w_{11}(x)=(-)\nonumber\\
\ \times\frac{\left|
\begin{array}{cccccccc}
c_1e^{\lambda_1}&\cdots&k_1^{n-2} c_1e^{\lambda_1} &k_1^{n} c_1e^{\lambda_1} &1&L_1&\cdots&L_1^{n-1}\\
c_2e^{\lambda_2}&\cdots&k_2^{n-2} c_2e^{\lambda_2} &k_2^{n} c_2e^{\lambda_2} &1&L_2&\cdots&L_2^{n-1}\\
\vdots&\vdots&\vdots&\vdots&\vdots&\vdots&\vdots&\vdots\\
c_{2n}e^{\lambda_{2n}}&\cdots&k_{2n}^{n-2} c_{2n}e^{\lambda_{2n}} &k_{2n}^{n} c_{2n}e^{\lambda_{2n}} &1&L_{2n}&\cdots&L_{2n}^{n-1}\\
\end{array}
\right|}{\left|
\begin{array}{cccccccc}
c_1e^{\lambda_1}&k_1 c_1e^{\lambda_1} &\cdots&k_1^{n-1} c_1e^{\lambda_1} &1&L_1&\cdots&L_1^{n-1}\\
c_2e^{\lambda_2}&k_2 c_2e^{\lambda_2} &\cdots&k_2^{n-1} c_2e^{\lambda_2} &1&L_2&\cdots&L_2^{n-1}\\
\vdots&\vdots&\vdots&\vdots&\vdots&\vdots&\vdots&\vdots\\
c_{2n}e^{\lambda_{2n}}&k_{2n} c_{2n}e^{\lambda_{2n}} &\cdots&k_{2n}^{n-1} c_{2n}e^{\lambda_{2n}} &1&L_{2n}&\cdots&L_{2n}^{n-1}\\
\end{array}
\right|}.\label{eq:w11}
\end{gather}
\end{subequations}
Thus, (\ref{eq:ourw}) is a solution to the differential--difference equation (\ref{eq:INLS1}) .\
Furthermore, (\ref{eq:ourw21}) will become a solution for (\ref{eq:rINLS}) if we impose certain conditions on $k_j$ and $c_j$.\ 
The following proposition constitutes the main result of this paper.\ 
\begin{proposition}\label{main}
{\rm
Suppose $k_j,\ a_j$ and $c_j$ satisfy the following conditions.\
\begin{itemize}
\item[C1]$\quad k_j^\dagger = -k_{j+n},\ a_j^\dagger =-a_{n+j}$ and $\xi$ is real.\
\item[C2]$\quad \ds\frac{\pi}{2\delta}>\textrm{Re }k_1>\textrm{Re }k_2>\cdots>\textrm{Re }k_n>0$
\item[C3]$\quad c_j,\ (j=1,2,\cdots,n)$ are given by
\begin{gather}
c_j=\prod_{r=1}^n\frac{L_j-L_{n+r}}{k_j-k_{n+r}},\qquad c_{n+j}=-\prod_{r=1\atop r\neq j}^n\frac{L_{n+j}-L_{n+r}}{k_{n+j}-k_{n+r}}.\nonumber
\end{gather}
\end{itemize}
Then, $w_{21}(x-i\delta)$ as in (\ref{eq:ourw21}) satisfies {\it the analyticity condition} and becomes a solution of (\ref{eq:rINLS}).\
}\end{proposition}

\vspace{.5\baselineskip}
{After imposing the above conditions, there still remain $4n$ real parameters 
in the determinants that appear in (\ref{eq:ourw}).\
The real and imaginary parts of the $k_j\ (j=1,2,\cdots,n)$ will define the amplitude and the velocity of each soliton.\
The dependence on the parameter $\xi$ can be absorbed in the real and imaginary parts of the $a_j\ (j=1,2,\cdots,n)$ which correspond to phase shifts.\
We show some examples of solitons before proving {\rm Proposition\ref{main}}.\
} 

\vspace{.5\baselineskip}
\begin{example}
The $1$--soliton solution is presented below.\ The parameters are set as $k_1=p +q i$, $k_2=-k_1^\dagger=-p +q i$ and $0<p <\ds\frac{\pi}{2\delta}$ in order to satisfy the conditions C1 and C2.\
$\xi$ is taken to be zero and $a_1=\alpha+\beta i$.\
\begin{gather}
w_{21}(x-i\delta)= \frac{\ds-\sqrt{p \sin 2\delta p }\ e^{i\{-q x+(p ^2-q ^2)t-\beta\}}}{\cosh\left[p (x-i\delta+2q t)+\delta q +\frac{1}{2}\log \frac{\sin2 \delta p }{p }+\alpha\right]}
\end{gather}
\end{example}

{ 
As for the $2$-soliton solution, the phase shift of the interaction can be explicitly calculated.\
Setting $k_j=p_j+iq_j$ and switching from the laboratory reference frame to the moving frame having the same velocity as the faster soliton whose velocity is equal to $-2q_1$ (for convenience we choose $q_1<q_2<0$,) (\ref{eq:ourw21}) is expressed approximately as
\begin{gather}
|w_{21}(x,t)|\sim 
\left\{
\begin{array}{ll}
\ds \cosh(p_1x)&(t\to-\infty)\\
\ds \cosh(p_1x+\theta )&(t\to\infty)
\end{array}
\right. ,
\end{gather} 
where the phase shift $\theta $ is given as follows.\
\begin{gather}
\ds e^{2\theta }=\frac{(p_1+p_2)^2+(q_1-q_2)^2}{(p_1-p_2)^2+(q_1-q_2)^2}\times \frac{\cosh2\delta (q_1-q_2)-\cos2\delta (p_1+p_2)}{\cosh2\delta (q_1-q_2)-\cos 2\delta (p_1-p_2)}.
\end{gather}
The phase shift for the slower soliton can be calculated in a similar manner, and one finds it to be exactly equal to $-\theta $.\ } 
%%%%%%%%%%%%%%%%%%%%%%%prop main%%%%%%%%%%%%%%%%%%%%%%%%%%%%%%%%%%%%%%%%%%%%%%%%%
\section{Proof of proposition \ref{main}} 
\quad Proposition \ref{main} will be proven in three steps.\
The first step is to show that $w_{21}(x-i\delta)$ and $w_{12}(x+i\delta)$ are complex conjugate.\
The second is to show that $w_{21}(x-i\delta)$ and $w_{12}(x+i\delta)$ vanish when $x$ goes to $\pm \infty$.\
The third and final step is to show that $w_{21}(x-i\delta)$ is analytic everywhere in the strip $0 \leq \textrm{Im }x\leq 2\delta$.\
Taken together the latter two statements constitute ``{\it the analyticity condition}'' of section 2.\
The proof will be carried out by reducing fractions and products made of differences.\
For simplicity some new notations are introduced.\

For an arbitrary set of subscripts $\sigma =\{\sigma_1,\sigma_2,\cdots,\sigma_a\}$ with the ordering $\sigma_1<\sigma_2<\cdots <\sigma_a$, we denote the Vandermonde determinant as
\begin{gather}
k_{\sigma}:=\prod_{1\leq j<j'\leq a}(k_{\sigma_j}-k_{\sigma_{j'}}),\qquad L_{\sigma}:=\prod_{1\leq j<j'\leq a}(L_{\sigma_j}-L_{\sigma_{j'}}).
\end{gather}
We also define a notation for products of differences between two disjoint sets of variables.\ 
Let $\mu=\{\mu_1,\mu_2,\cdots,\mu_{b}\}$ be a set of subscripts with the ordering $\mu_1<\mu_2<\cdots <\mu_b$, $\sigma$ as above, then we define,
\begin{gather}
k_{\sigma\to\mu}:=\prod_{j=1}^{a}\prod_{j'=1}^{b}(k_{\sigma_j}-k_{\mu_{j'}})\, ,\qquad L_{\sigma\to\mu}:=\prod_{j=1}^{a}\prod_{j'=1}^{b}(L_{\sigma_j}-L_{\mu_{j'}})\, .
\end{gather}

The following statements will be used extensively.\

\begin{remark}%%%%%REMARK1%%%%%%%%%%%%%%%%%%%%%%%%%%%  
{ 
According to the condition C1, $k_j^\dagger=-k_{n+j},\ L_j^\dagger=-L_{n+j}$
and $\lambda_j^\dagger=-\lambda_{n+j}$ for real $x, t$.\
In these equalities, the subscripts should be read modulo $2n$.\
}
\end{remark}
\begin{remark} 
The real parts of the $L_j$'s are positive for $j=1,2,\cdots,n$ due to condition C2.\
\end{remark}
\begin{remark} 
Let $\sigma:=\{\sigma_1,\sigma_2,\cdots,\sigma_a\}$ be a set of subscripts and consider products of differences $k_\sigma$ and $L_\sigma$.\ If we split $\sigma$ into two arbitrary disjoint subsets $\mu:=\{\mu_1,\mu_2,\cdots,\mu_{b}\}$ , $\kappa:=\{\kappa_1,\kappa_2,\cdots,\kappa_{c}\}$ ($b+c=a$), we have the decompositions:
\begin{gather}
\frac{L_\sigma}{k_\sigma}=\frac{L_\mu L_\kappa L_{\mu\to\kappa}}{k_\mu k_\kappa k_{\mu\to\kappa}},\quad L_\sigma k_\sigma =L_\mu L_\kappa L_{\mu\to\kappa}k_\mu k_\kappa k_{\mu\to\kappa}.\ \label{eq:diffprodeform1}
\end{gather}  
Note that the equalities do not hold for $L_\sigma$ or $k_\sigma$ independently, i.e.,
\begin{gather}
L_\sigma \neq L_\mu L_\kappa L_{\mu\to\kappa}\ ,\quad 
k_\sigma \neq k_\mu k_\kappa k_{\mu\to\kappa}\ ,
\end{gather}
as the signs can differ on both sides of the relations due to the ordering of the subscripts.\
Obviously these signs are cancelled when we think of a fraction or product like (\ref{eq:diffprodeform1}).\ 
\end{remark}
Now we proceed to the proof of proposition {\ref {main}}.\ 
A first result we need is the following theorem, 
which shall be proven in the Appendix.\
%%%%%%%%%%%%%%%%%%%%%%Conjugation%%%%%%%%%%%%%%%%%%%%%%%%%%%%%%%%%%%%
\begin{theorem}\label{conjugation}
$w_{21}$ and $w_{12}$ are complex conjugate.\
\end{theorem}
%%%%%%%%%%%%%%%%%%%%%%%%%%%%%%%%%%%%%%%%%%%%%%%%%%%%%%%%%%%%%%%%%%%%
\vspace{2mm}

Secondly we have

\begin{property}\label{steepness}
$w_{21}(x-i\delta)$ and $w_{12}(x+i\delta)$ go to $0$ when $x$ goes to $\pm \infty$.\
\end{property}
\begin{proof}

Because of condition C2 in Proposition \ref{main}, we see that the first $n-1\ (n)$ rows become dominant in the numerator (denominator) of (\ref{eq:ourw21}) when $x$ goes to $\infty$.\ Thus, we have
\begin{gather}
\begin{array}{ll}
\ds w_{21}(x)\stackrel{x\to\infty}{\longrightarrow}&\ds (-)^{n+1}\frac{\ds k_{\{1,\cdots,n-1\}}L_{\{n,\cdots,2n\}}\prod_{r=1}^{n-1}e^{\lambda_r}}{\ds k_{\{1,\cdots,n\}}L_{\{n+1,\cdots,2n\}}\prod_{r=1}^{n}e^{\lambda_r}}\\
&\ds =
(-)^{n+1}\frac{\ds L_{\{n\}\to\{n+1,\cdots,2n\}}}{\ds k_{\{1,\cdots,n-1\}\to\{n\}}}e^{-\lambda_n}\to 0
\end{array}
\label{eq:xbecomslarger}
\end{gather}
On the contrary, the rows from the $n+1$th to the $2n-1$th (the $2n$th) become dominant when $x$ goes to $-\infty$, which implies,  
\begin{gather}
\begin{array}{ll}
\ds w_{21}(x)\stackrel{x\to -\infty}{\longrightarrow}&\ds (-)\frac{\ds k_{\{n+1,\cdots,2n-1\}}L_{\{1,\cdots,n,2n\}}\prod_{r=n+1}^{2n-1}e^{\lambda_r}}{\ds k_{\{n+1,\cdots,2n\}}L_{\{1,\cdots,n\}}\prod_{r=1}^{n}e^{\lambda_r}}\\
&\ds =\frac{\ds -L_{\{1,\cdots,n\}\to\{2n\}}}{\ds k_{\{n+1,\cdots,2n-1\}\to\{2n\}}}e^{-\lambda_{2n}}\to 0
\end{array}
\label{eq:xbecomessmaller}
\end{gather}
This procedure also applies to $w_{21}(x-i\delta)$ and $w_{12}(x+i\delta)$.\
\end{proof}

%%%%%%%%%%%%%%%%%%%%%%%%%%%%%%%%Cauchy determinant%%%%%%%%%%%%%%%%%%%%%%%%%%
\hspace{3mm}

To accomplish the third step of the proof, we first change the denominator of $w_{21}$ to a determinant of the sum of two matrices of half the size.\ 
\begin{theorem}\label{thmanalyticity}%%%%%%%%%%%%thmanalyticity%%%%%%%%%%%
\begin{gather}
\begin{array}{l}
\ds \frac{k_{\{n+1\cdots 2n\}}}{L_{\{n+1\cdots 2n\}}}
\times\frac{1}{L_{\{1,\cdots,n\}\to\{n+1,\cdots,2n\}}}
\times \tau\\[5mm]
\ds \quad =(-)^{\frac{n(n-1)}{2}}{\rm det}\left(\frac{e^{\lambda_{n+j}}}{L_i-L_{n+j}}+\frac{e^{\lambda_i}}{k_i-k_{n+j}}\right)_{1\leq i,j \leq n},
\end{array}
\label{eq:analyticity1}
\end{gather}
where $\tau$ denotes the denominator of (\ref{eq:ourw21}).\ 
\end{theorem}
\begin{proof}

First we express the right hand side of (\ref{eq:analyticity1}) as a sum, each term of which is the product of a minor determinant of the matrix $\ds \left(\frac{e^{\lambda_i}}{k_i-k_{n+j}}\right)_{ij}$ and a minor of the matrix $\ds\left(\frac{e^{\lambda_{n+j}}}{L_i-L_{n+j}}\right)_{ij}$.\
Thus, we have   
\begin{gather}
\begin{array}{l}
\ds (-)^{\frac{n(n-1)}{2}}{\rm det}\left(\frac{e^{\lambda_{n+j}}}{L_i-L_{n+j}}+\frac{e^{\lambda_i}}{k_i-k_{n+j}}\right)\\
\ds \quad =(-)^{\frac{n(n-1)}{2}}\sum_{\{\sigma,\sigma'\}\atop \{\mu,\mu'\}}(-)^{[\sigma]+[\mu]}{\rm det}\left(\frac{1}{k_{\sigma_i}-k_{n+\mu_j}}\right){\rm det}\left(\frac{1}{L_{\sigma'_i}-L_{n+\mu'_j}}\right).
\end{array}\label{eq:cauchy1}
\end{gather}
More precisely, when expanding the left hand side we first choose rows, from which we then choose the elements of matrix $\ds\left(\frac{e^{\lambda_i}}{k_i-k_{n+j}}\right)_{ij}$.\ 
We label these rows $\sigma:=\{\sigma_1,\sigma_2,\cdots,\sigma_r\}$.\  
The remaining rows are labeled $\sigma':=\{\sigma'_1,\sigma'_2,\cdots,\sigma'_{s}\}$, which yield the the elements of $\ds\left(\frac{e^{\lambda_{n+j}}}{L_i-L_{n+j}}\right)_{ij}$.\ 

We also have to decide on a partition of the columns.\ 
We label the columns belonging to $\ds\left(\frac{e^{\lambda_i}}{k_i-k_{n+j}}\right)_{ij}$ as $\mu:=\{\mu_1,\mu_2,\cdots,\mu_r\}$ (the number of elements in $\sigma$ and $\mu$ must of course be the same).\ 
The remaining columns are denoted as $\mu':=\{\mu'_1,\cdots,\mu'_s\}$.\ 

It should be noted that we keep the internal ordering of these subsets to be ascending.\ The summation that appears in (\ref{eq:cauchy1}) runs over all possible partitions $\{\sigma,\sigma'\}$ and $\{\mu,\mu'\}$.\ 
More detail regarding the decomposition of the determinant of a sum of two matrices into minor determinants can be found in ~\cite{f}.\  

Since we will use the subscripts $\{n+\mu_1,\cdots,n+\mu_r\}$ and $\{n+\mu'_{1},\cdots,n+\mu'_{s}\}$ more often than $\mu$ and $\mu'$ themselves, we denote them as $n+\mu$ and $n+\mu'$.\ 
Now, each term in (\ref{eq:cauchy1}) can be recognized to be a Cauchy determinant.\ 
So we can express (\ref{eq:cauchy1}) as
\begin{gather}
\begin{array}{l}
\ds (-)^{\frac{n(n-1)}{2}}{\rm det}\left(\frac{e^{\lambda_{n+j}}}{L_i-L_{n+j}}+\frac{e^{\lambda_i}}{k_i-k_{n+j}}\right)\\[4mm]
\ds \quad =(-)^{\frac{n(n-1)}{2}}\sum_{\{\sigma,\sigma'\}\atop \{\mu,\mu'\}}(-)^{[\sigma]+[\mu]+\frac{r(r-1)}{2}+\frac{s(s-1)}{2}}\\[4mm]
\ds \hspace{3cm}\times\frac{k_\sigma k_{n+\mu}L_{\sigma'}L_{n+\mu'}}{k_{\sigma\to n+\mu}L_{\sigma'\to n+\mu'}}e^{\sum_{j=1}^r\lambda_{\sigma_j}+\sum_{j=1}^{s}\lambda_{n+\mu'_{j}}}\\[4mm]
\ds \quad =\sum_{\{\sigma,\sigma'\}\atop \{\mu,\mu'\}}(-)^{[\sigma]+[\mu]+(n+1)r}\frac{k_\sigma k_{n+\mu}L_{\sigma'}L_{n+\mu'}}{k_{\sigma\to n+\mu}L_{\sigma'\to n+\mu'}}e^{\sum_{j=1}^r\lambda_{\sigma_j}+\sum_{j=1}^{s}\lambda_{n+\mu'_{j}}}
\end{array}\label{eq:cauchy2}
\end{gather}
Next we expand the l.h.s. of (\ref{eq:analyticity1}) and verify that the coefficient in $e^{\sum_{j=1}^r\lambda_{\sigma_j}+\sum_{j=1}^{s}\lambda_{n+\mu'_j}}$ is the same as that of (\ref{eq:cauchy2}).\
The expansion involves a rather cumbersome parity which we temporarily denote as $(-)^S$.\ 
Due to Remark3 we have the following.\
\begin{gather}
\ds \textrm{the coefficient of }e^{\sum_{j=1}^r\lambda_{\sigma_j}+\sum_{j=1}^{s}\lambda_{n+\mu'_j}}\textrm{ in the l.h.s. of (\ref{eq:analyticity1})}\notag\\
\ds =\frac{k_{\{n+1,\cdots,2n\}}}{L_{\{n+1,\cdots,2n\}}}\frac{1}{L_{\{1,\cdots,n\}\to\{n+1,\cdots,2n\}}}\times\prod_{j=1}^rc_{\sigma_j}\times\prod_{j=1}^sc_{n+\mu'_j}\notag\\
\ds \qquad\qquad \times \left|
\begin{array}{cccc}
1&k_{\sigma_1}&\cdots&k_{\sigma_1}^{n-1}\\
\vdots&\vdots&\vdots&\vdots\\
1&k_{\sigma_r}&\cdots&k_{\sigma_r}^{n-1}\\
1&k_{n+\mu'_1}&\cdots&k_{n+\mu'_1}^{n-1}\\
\vdots&\vdots&\vdots&\vdots\\
1&k_{n+\mu'_s}&\cdots&k_{n+\mu'_s}^{n-1}\\
\end{array}
\right|
\left|
\begin{array}{cccc}
1&L_{\sigma'_1}&\cdots&L_{\sigma'_1}^{n-1}\\
\vdots&\vdots&\vdots&\vdots\\
1&L_{\sigma'_s}&\cdots&L_{\sigma'_s}^{n-1}\\
1&L_{n+\mu_1}&\cdots&L_{n+\mu_1}^{n-1}\\
\vdots&\vdots&\vdots&\vdots\\
1&L_{n+\mu_r}&\cdots&L_{n+\mu_r}^{n-1}\\
\end{array}
\right|\times(-)^{S}\notag
\end{gather}%%end of the determinant%%
\begin{gather}
\ds =(-)^{S}\times\frac{k_{\{n+1,\cdots,2n\}}}{L_{\{n+1,\cdots,2n\}}}\frac{k_{\{\sigma,n+\mu'\}}L_{\{\sigma',n+\mu\}}}{L_{\{1,\cdots,n\}\to\{n+1,\cdots,2n\}}}\prod_{j=1}^rc_{\sigma_j}\times\prod_{j=1}^sc_{n+\mu'_j}\notag\\
\ds =(-)^{S}\times\frac{k_{\{n+1,\cdots,2n\}}}{L_{\{n+1,\cdots,2n\}}}\frac{k_\sigma k_{n+\mu'} k_{\sigma\to n+\mu'} L_{\sigma'}L_{n+\mu}L_{\sigma'\to n+\mu}}{L_{\{1,\cdots,n\}\to\{n+1,\cdots, 2n\}}}\notag\\
\ds\qquad\qquad \times\prod_{j=1}^r\frac{L_{\sigma_j\to\{n+1,\cdots,2n\}}}{k_{\sigma_j\to \{n+1,\cdots,2n\}}}\times
\left(\prod_{j=1}^s-\frac{L_{n+\mu'_j\to \{n+1,\cdots,2n\}\backslash\{n+\mu'_j\}}}{k_{n+\mu'_j\to \{n+1,\cdots,2n\}\backslash\{n+\mu'_j\}}}\right)\notag\\%%
\ds =(-)^{S}\times (-)^s\times\frac{k_{\{n+1,\cdots,2n\}}}{L_{\{n+1,\cdots,2n\}}}\frac{k_\sigma k_{n+\mu'} k_{\sigma\to n+\mu'} L_{\sigma'}L_{n+\mu}L_{\sigma'\to n+\mu}}{L_{\{1,\cdots,n\}\to\{n+1,\cdots,2n\}}}\notag\\
\ds \qquad\qquad \times\frac{L_{\sigma\to\{n+1,\cdots,2n\}}}{k_{\sigma\to\{n+1,\cdots,2n\}}}\left(\frac{L_{n+\mu'}}{k_{n+\mu'}}\right)^2\frac{L_{n+\mu'\to n+\mu}}{k_{n+\mu'\to n+\mu}}\notag\\
\ds =(-)^{S}\times (-)^s\times\frac{k_\sigma k_{n+\mu}L_{\sigma'}L_{n+\mu'}}{k_{\sigma\to n+\mu}L_{\sigma'\to n+\mu'}}
\label{eq:cauchy3}
\end{gather}
Thus, the remaining task is to show that the signs of (\ref{eq:cauchy2}) and (\ref{eq:cauchy3}) are the same.\ 
In (\ref{eq:cauchy3}), $(-)^{S}$ denotes the parity of the permutation needed to rearrange $\{1,2,\cdots,2n\}$ in the order $\{\sigma,n+\mu',\sigma',n+\mu\}$.\ In this rearrangement, we first change $\{1,2,\cdots,2n\}$ to $\{\sigma,\sigma',n+\mu,n+\mu'\}$, which has the parity $(-)^{[\sigma]+[\mu]}$.\ 
Then we change $\{\sigma,\sigma',n+\mu,n+\mu'\}$ to $\{\sigma,n+\mu',\sigma',n+\mu\}$, which has the parity $(-)^{ns}$.\ 
Thus, we have
\begin{gather}
(-)^S\times (-)^s=(-)^{[\sigma]+[\mu]+(n+1)s}=(-)^{[\sigma]+[\mu]+(n+1)r}.\label{eq:cauchy4} 
\end{gather}
From (\ref{eq:cauchy2}), (\ref{eq:cauchy3}), (\ref{eq:cauchy4}), We see that {\rmfamily Theorem \ref{thmanalyticity}} holds.\ 
\end{proof}
%%%%%%%%%%%%%%%%%%%%%%%%%%%%%%%analyticity%%%%%%%%%%%%%%%%%%%%%%

Using Theorem \ref{thmanalyticity}, we can now accomplish the last step in the proof of Proposition\ref{main}
\begin{corollary}
 If the absolute values of $k_j$ are small enough, $w_{21}(x-i\delta)$ becomes analytic everywhere in the strip $0\leq {\rm Im}\; x \leq 2\delta$.\
\end{corollary}
\begin{proof} 

We shall prove that $w_{21}(x)$ is analytic everywhere in $\delta\geq |{\rm Im}\; x|$.\
We first rewrite {\rmfamily Theorem \ref{thmanalyticity}} as follows.\
\begin{gather}
\begin{array}{ll}
\tau
\ds =(-)^{\frac{n(n-1)}{2}}
\times \frac{L_{\{n+1\cdots 2n\}}}{k_{\{n+1\cdots 2n\}}}
&\ds \times L_{\{1,\cdots,n\}\to\{n+1,\cdots,2n\}}
\times e^{\sum_{J=1}^n\lambda_{n+j}}\\
&\ds \times{\rm det}\left(\frac{1}{L_i-L_{n+j}}+\frac{e^{\lambda_i-\lambda_{n+j}}}{k_i-k_{n+j}}\right).
\end{array}\label{eq:lastcor}
\end{gather}
So, whether $w_{21}(x)$ is analytic or not depends on the last determinant in (\ref{eq:lastcor}).\ 
Due to Remark 1, the determinant becomes
\begin{gather}
{\rm det}\left(\frac{1}{L_i+L_{j}^\dagger}+\frac{e^{\lambda_i+\lambda_{j}^\dagger}}{k_i+k_{j}^\dagger}\right). \label{eq:cauchy5}
\end{gather} 
It is easily seen that (\ref{eq:cauchy5}) is positive for real $x$ because, for any nonzero vector $(y_1,y_2,\cdots,y_n)$,  
\begin{gather}
\begin{array}{l}
\ds (y_1,\cdots,y_n)
\left(
\begin{array}{ccc}
\frac{1}{L_1+L_{1}^\dagger}+\frac{e^{\lambda_1+\lambda_{1}^\dagger}}{k_1+k_{1}^\dagger}&\cdots&\frac{1}{L_1+L_{n}^\dagger}+\frac{e^{\lambda_1+\lambda_{n}^\dagger}}{k_1+k_{n}^\dagger}\\
\vdots&\ddots&\vdots\\
\frac{1}{L_n+L_{1}^\dagger}+\frac{e^{\lambda_n+\lambda_{1}^\dagger}}{k_n+k_{1}^\dagger}&\cdots&\frac{1}{L_n+L_{n}^\dagger}+\frac{e^{\lambda_n+\lambda_{n}^\dagger}}{k_i+k_{j}^\dagger}
\end{array}
\right)
\left(
\begin{array}{c}
y_1^\dagger\\
\vdots\\
y_n^\dagger
\end{array}
\right)\\
\ds =\int_{-\infty}^0du
(y_1,\cdots,y_n)
\left(
\begin{array}{ccc}
e^{(L_1+L_{1}^\dagger)u}&\cdots&e^{(L_1+L_{n}^\dagger)u}\\
\vdots&\ddots&\vdots\\
e^{(L_n+L_{1}^\dagger)u}&\cdots&e^{(L_n+L_{n}^\dagger)u}
\end{array}
\right)
\left(
\begin{array}{c}
y_1^\dagger\\
\vdots\\
y_n^\dagger
\end{array}
\right)\\
\ds \qquad +\int_{-\infty}^xdu
(y_1,\cdots,y_n)
\left(
\begin{array}{ccc}
e^{\lambda_1+\lambda_{1}^\dagger}|_{x=u}&\cdots&e^{\lambda_1+\lambda_{n}^\dagger}|_{x=u}\\
\vdots&\ddots&\vdots\\
e^{\lambda_n+\lambda_{1}^\dagger}|_{x=u}&\cdots&e^{\lambda_n+\lambda_{n}^\dagger}|_{x=u}
\end{array}
\right)
\left(
\begin{array}{c}
y_1^\dagger\\
\vdots\\
y_n^\dagger
\end{array}
\right)\\
\ds =\int_{-\infty}^0du\sum_{j=1}^n|y_je^{L_ju}|^2+\int_{-\infty}^xdu\sum_{j=1}^n|(y_je^{\lambda_j}|_{x=u})|^2>0.
\end{array}
\end{gather}
Regarding the integrals aboves, note that $k_j+k_j^\dagger$ and $L_j+L_j^\dagger$ are positive due to condition C1 and Remark2.\ 
Thus, we see (\ref{eq:cauchy5}) is nonzero on the real axis.\ 
As for the whole strip $\delta\leq |{\rm Im}\; x|$, by inspection of (\ref{eq:xbecomslarger}) and (\ref{eq:xbecomessmaller}), we see that $w_{21}$ is analytic in the strip if the real part of $|x|$ is large enough.\ 
More precisely, there exists a real constant $R$ such that $w_{21}$ is analytic in the region $\ds\{x\left|\delta\leq |{\rm Im}\; x| \cap |{\rm Re}\;x|\geq R \right.\}$,
which leaves open the possibility that $w_{21}$ might still have singular points in the rectangle $\ds\{x\left|\delta\leq |{\rm Im}\; x| \cap |{\rm Re }x|< R \right.\}$.\ 
However, since $w_{21}$ is real analytic, we have a certain analytic neighbourhood including the line $\{x|-R\leq x\leq R\}$.\ 
Furthermore, since $w_{21}$ contains the variable $x$ only in the form $k_jx$, choosing the absolute values of $k_j$ small enough, one can make sure the analytic neighbourhood gets broad enough until it contains the whole rectangle.\ 
\end{proof}
%%%%%%%%%%%%%%%%%%%%%%%%%%%%%%%%%Concluding remarks%%%%%%%%%%%%%%%%%
\section{Concluding remarks}
\quad We presented a double Wronskian solution for the focusing intermediate nonlinear Schr\"{o}dinger equation.\ 
Its construction was based on Sato theory, which also clarified the relation between the $2$--component KP hierarchy and this particular INLS equation.\

The solitons exhibit peculiar interactions, i.e.,  
the $2$--soliton solution exhibits oscillations for the slower soliton in the interaction region.\
 It is an interesting problem to investigate the properties of these soliton solutions in comparison to those of e.g. the NLS equation, which should shed more light on the characteristic interaction properties of the INLS solitons.\
This will be investigated in the future.\ 

{ 
As pointed out in the Introduction, the $\delta\to0$ limit of the INLS equation with $\sigma=-1$ yields the focusing NLS equation.\
It is important to remark that the solution (\ref{eq:ourw21}) satisfying $C1\sim C3$ endures this limit, i.e. $U=w_{21}/\sqrt{2\delta}$ corresponds to the $n$-soliton solution of the focusing NLS in double Wronskian form.\
}
Let us explain the fact briefly.\
Since $c_j\to (2\delta)^n, c_{n+j}\to -(2\delta)^{n-1},\  (j=1,2,\cdots,n)$ and $L_j\to 2\delta k_j,\ (j=1,2,\cdots,2n)$ as $\delta\to0$, $w_{21}$ can be approximated as follows.\  
\begin{gather}
w_{21}(x)/\sqrt{2\delta}=(-)^{n+1}\nonumber\\
\ \times\frac{ 
\left|
\begin{array}{ccccccc}
\sqrt{2\delta} e^{\lambda_1}&\cdots&k_1^{n-2} \sqrt{2\delta}e^{\lambda_1} &1&k_1&\cdots&k_1^{n}\\
\sqrt{2\delta} e^{\lambda_2}&\cdots&k_2^{n-2} \sqrt{2\delta}e^{\lambda_2} &1&k_2&\cdots&k_2^{n}\\
\vdots&\vdots&\vdots&\vdots&\vdots&\vdots&\vdots\\
-(\sqrt{2\delta})^{-1}e^{\lambda_{2n}}&\cdots&-k_{2n}^{n-2} (\sqrt{2\delta})^{-1}e^{\lambda_{2n}} &1&k_{2n}&\cdots&k_{2n}^{n}\\
\end{array}
\right|}{
\left|
\begin{array}{ccccccc}
\sqrt{2\delta} e^{\lambda_1}&\cdots&k_1^{n-1} \sqrt{2\delta}e^{\lambda_1} &1&k_1&\cdots&k_1^{n-1}\\
\sqrt{2\delta} e^{\lambda_2}&\cdots&k_2^{n-1} \sqrt{2\delta}e^{\lambda_2} &1&k_2&\cdots&k_2^{n-1}\\
\vdots&\vdots&\vdots&\vdots&\vdots&\vdots&\vdots\\
-(\sqrt{2\delta})^{-1}e^{\lambda_{2n}}&\cdots&-k_{2n}^{n-1} (\sqrt{2\delta})^{-1}e^{\lambda_{2n}} &1&k_{2n}&\cdots&k_{2n}^{n-1}\\
\end{array}
\right|}
\label{eq:NLS}
\end{gather}
By defining the phase constants $a_j=a'_j-\frac12\ln(2\delta)$, the $\sqrt{2\delta}$'s in the determinants disappear and (\ref{eq:NLS}) becomes nothing else but the bright $n$-soliton solution of the focusing NLS equation.\    

%%%%%%%%%%%%%%%%%%%%%%%%%

\vspace{\baselineskip}
\noindent{\bf Acknowledgement}

This work was partially supported by a Grant-in-Aid from the Japan foundation for the Promotion of Science (JSPS).\ 
The author is grateful to R. Willox for his pointed remarks and for encouraging the author.\ 
The author also would like to thank S. Kakei, Y. Ohta and J. Nimmo for discussions and advice.\
This paper could not have been refined without their help.\ 
%%%%%%%%%%%%%%%%%%Proof of Theorem \ref{conjugation}%%%%%%%%%%%%%%%%%%%%
\appendix
\section{Proof of Theorem \ref{conjugation}}
\begin{proof}
By virtue of Remark 1, the complex conjugate of (\ref{eq:ourw21}) is   
\begin{gather}
w_{21}^\dagger=(-)^{n+1}\nonumber\\
\times
\frac{\left|
\begin{array}{cccccc}
c_1^\dagger e^{-\lambda_{n+1}}&\cdots&(-k_{n+1})^{n-2} c_1^\dagger e^{-\lambda_{n+1}} &1&\cdots&(-L_{n+1})^{n}\\
\vdots&\vdots&\vdots&\vdots&\vdots&\vdots\\
c_n^\dagger e^{-\lambda_{2n}}&\cdots&(-k_{2n})^{n-2} c_n^\dagger e^{-\lambda_{2n}} &1&\cdots&(-L_{2n})^{n}\\
c_{n+1}^\dagger e^{-\lambda_1}&\cdots&(-k_1)^{n-2} c_{n+1}^\dagger e^{-\lambda_1} &1&\cdots&(-L_1)^{n}\\
\vdots&\vdots&\vdots&\vdots&\vdots&\vdots\\
c_{2n}^\dagger e^{-\lambda_{n}}&\cdots&(-k_{n})^{n-2} c_{2n}^\dagger e^{-\lambda_{n}} &1&\cdots&(-L_{n})^{n}\\
\end{array}
\right|\;} 
{\left|
\begin{array}{cccccc}
c_1^\dagger e^{-\lambda_{n+1}}&\cdots&(-k_{n+1})^{n-1} c_1^\dagger e^{-\lambda_{n+1}} &1&\cdots&(-L_{n+1})^{n-1}\\
\vdots&\vdots&\vdots&\vdots&\vdots&\vdots\\
c_n^\dagger e^{-\lambda_{2n}}&\cdots&(-k_{2n})^{n-1} c_n^\dagger e^{-\lambda_{2n}} &1&\cdots&(-L_{2n})^{n-1}\\
c_{n+1}^\dagger e^{-\lambda_1}&\cdots&(-k_1)^{n-1} c_{n+1}^\dagger e^{-\lambda_1} &1&\cdots&(-L_1)^{n-1}\\
\vdots&\vdots&\vdots&\vdots&\vdots&\vdots\\
c_{2n}^\dagger e^{-\lambda_{n}}&\cdots&(-k_{n})^{n-1} c_{2n}^\dagger e^{-\lambda_{n}} &1&\cdots&(-L_{n})^{n-1}\\
\end{array}
\right|}.\label{eq:conj1}%%%%complex conjugate%%%%%%%%%% 
\end{gather}
We arrange the rows in ascending order for the subscripts of $\lambda$ and gather the negative signs in each row.\ Then we have,
\begin{gather}
w_{21}^\dagger=
\frac{(-)^{n}\left|
\begin{array}{cccccc}
c_{n+1}^\dagger e^{-\lambda_1}&\cdots&k_1^{n-2} c_{n+1}^\dagger e^{-\lambda_1} &1&\cdots&L_1^{n}\\
\vdots&\vdots&\vdots&\vdots&\vdots&\vdots\\
c_{2n}^\dagger e^{-\lambda_{n}}&\cdots&k_{n}^{n-2} c_{2n}^\dagger e^{-\lambda_{n}} &1&\cdots&L_{n}^{n}\\
c_1^\dagger e^{-\lambda_{n+1}}&\cdots&k_{n+1}^{n-2} c_1^\dagger e^{-\lambda_{n+1}} &1&\cdots&L_{n+1}^{n}\\
\vdots&\vdots&\vdots&\vdots&\vdots&\vdots\\
c_n^\dagger e^{-\lambda_{2n}}&\cdots&k_{2n}^{n-2} c_n^\dagger e^{-\lambda_{2n}} &1&\cdots&L_{2n}^{n}\\
\end{array}
\right|\;} 
{\left|
\begin{array}{cccccccc}
c_{n+1}^\dagger e^{-\lambda_1}&\cdots&k_1^{n-1} c_{n+1}^\dagger e^{-\lambda_1} &1&\cdots&L_1^{n-1}\\
\vdots&\vdots&\vdots&\vdots&\vdots&\vdots\\
c_{2n}^\dagger e^{-\lambda_{n}}&\cdots&k_{n}^{n-1} c_{2n}^\dagger e^{-\lambda_{n}} &1&\cdots&L_{n}^{n-1}\\
c_1^\dagger e^{-\lambda_{n+1}}&\cdots&k_{n+1}^{n-1} c_1^\dagger e^{-\lambda_{n+1}} &1&\cdots&L_{n+1}^{n-1}\\
\vdots&\vdots&\vdots&\vdots&\vdots&\vdots\\
c_n^\dagger e^{-\lambda_{2n}}&\cdots&k_{2n}^{n-1} c_n^\dagger e^{-\lambda_{2n}} &1&\cdots&L_{2n}^{n-1}\\
\end{array}
\right|}.\label{eq:conj2}%%%complex conjugate%%%
\end{gather}
Now we carry out the Laplace expansions of both the numerator and the denominator in (\ref{eq:conj2}).\
\begin{gather}
\begin{array}{l}
\ds \textrm{The numerator of (\ref{eq:conj2})}\\
\ds \quad =(-)^{n}\times 
\displaystyle\sum(-)^{[\sigma]}\prod_{r=1}^{n-1}c_{n+\sigma_r}^\dagger e^{-\lambda_{\sigma_r}}\\
\ds \qquad\qquad \times\left|
\begin{array}{cccc}
1&k_{\sigma_1}&\cdots&k_{\sigma_1}^{n-2}\\
1&k_{\sigma_2}&\cdots&k_{\sigma_2}^{n-2}\\
\vdots&\vdots&\vdots&\vdots\\
1&k_{\sigma_{n-1}}&\cdots&k_{\sigma_{n-1}}^{n-2}
\end{array}
\right|
\left|
\begin{array}{cccc}
1&L_{\sigma'_1}&\cdots&L_{\sigma'_1}^{n}\\
1&L_{\sigma'_2}&\cdots&L_{\sigma'_2}^{n}\\
\vdots&\vdots&\vdots&\vdots\\
1&L_{\sigma'_{n+1}}&\cdots&L_{\sigma'_{n+1}}^{n}
\end{array}
\right|\\
\ds \quad =(-)^{n+1}\times 
\displaystyle\sum(-)^{[\sigma]}k_\sigma L_{\sigma'}\prod_{r=1}^{n-1}c_{n+\sigma_r}^\dagger e^{-\lambda_{\sigma_r}}\end{array}
\end{gather}
The summation runs over all possible ascending partitions $\sigma,\sigma'$ of $\{1,2,\cdots,2n\}$, where $\sigma:=\{\sigma_1,\sigma_2,\cdots,\sigma_{n-1}\}$ and $\sigma':=\{\sigma'_1,\sigma'_2,\cdots,\sigma'_{n+1}\}$.\ 
$(-)^{[\sigma]}$ denotes the parity of the permutation needed to move the rows of type $\sigma$ to the beginning and of type $\sigma'$ to the end.\ 
We apply the same procedure to the denominator of (\ref{eq:conj2}) (for which the partition is denoted by $\mu:=\{\mu_1,\mu_2,\cdots,\mu_{n}\},\mu':=\{\mu'_1,\mu'_2,\cdots,\mu'_{n}\}$) and have
\begin{gather}
w_{21}^\dagger=\frac{
(-)^{n+1}\displaystyle\sum_{\{\sigma_j\},\{\sigma'_j\}}(-)^{[\sigma]}k_\sigma L_{\sigma'}\prod_{r=1}^{n-1}c_{n+\sigma_r}^\dagger e^{-\lambda_{\sigma_r}}
}
{
\displaystyle\sum_{\mu,\mu'}(-)^{[\mu]}k_\mu L_{\mu'}\prod_{r=1}^{n}c_{n+\mu_r}^\dagger e^{-\lambda_{\mu_r}}
}.\label{eq:w21dagger}
\end{gather}
Next we consider $w_{12}$.\ 
Rearranging the columns in (\ref{eq:w12}), we have,
\begin{gather}
\begin{array}{ll}
\ds w_{12}&=\frac{(-)^{n+1}\left|
\begin{array}{cccccc}
c_1^{-1}e^{-\lambda_1}&\cdots&L_1^{n-2} c_1^{-1}e^{-\lambda_1}&1&\cdots&k_1^{n}\\
c_2^{-1}e^{-\lambda_2}&\cdots&L_2^{n-2} c_2^{-1}e^{-\lambda_2}&1&\cdots&k_2^{n} \\
\vdots&\vdots&\vdots&\vdots&\vdots&\vdots\\
c_{2n}^{-1}e^{-\lambda_{2n}}&\cdots&L_{2n}^{n-2} c_{2n}^{-1}e^{-\lambda_{2n}}&1&\cdots&k_{2n}^{n}\\
\end{array}
\right|}{\left|
\begin{array}{cccccccc}
c_1^{-1}e^{-\lambda_1}&\cdots&L_1^{n-1} c_1^{-1}e^{-\lambda_1}&1&\cdots&k_1^{n-1}\\
c_2^{-1}e^{-\lambda_2}&\cdots&L_2^{n-1} c_2^{-1}e^{-\lambda_2}&1&\cdots&k_2^{n-1}\\
\vdots&\vdots&\vdots&\vdots&\vdots&\vdots\\
c_{2n}^{-1}e^{-\lambda_{2n}}&\cdots&L_{2n}^{n-1} c_{2n}^{-1}e^{-\lambda_{2n}}&1&\cdots&k_{2n}^{n-1}\\
\end{array}
\right|}\\
&\ds =\frac{
(-)^{n}\displaystyle\sum_{\sigma,\sigma'}(-)^{[\sigma]}L_\sigma k_{\sigma'}\prod_{r=1}^{n-1}c_{\sigma_r}^{-1} e^{-\lambda_{\sigma_r}}
}
{
\displaystyle\sum_{\mu,\mu'}(-)^{[\mu]}L_\mu k_{\mu'}\prod_{r=1}^{n}c_{\mu_r}^{-1} e^{-\lambda_{\mu_r}}
}.
\end{array}\label{eq:w12dagger}
\end{gather}
To prove $w_{21}^\dagger=w_{12}$ all we have to do is to show that 
\begin{gather}
\begin{array}{l}
\ds\frac{\textrm{the coefficient of }e^{-\sum_{j=1}^{n-1}\lambda_{\sigma_j}}\textrm{of the numerator of }(\ref{eq:w21dagger})}{\textrm{the coefficient of }e^{-\sum_{j=1}^{n-1}\lambda_{\sigma_j}}\textrm{of the numerator of }(\ref{eq:w12dagger})}\\
\ds\qquad =\frac{\textrm{the coefficient of }e^{-\sum_{j=1}^n\lambda_{\mu_j}}\textrm{of the denominator of }(\ref{eq:w21dagger})}{\textrm{the coefficient of }e^{-\sum_{j=1}^n\lambda_{\mu_j}}\textrm{of the denominator of }(\ref{eq:w12dagger})}
\end{array}\label{eq:coefficientconjugation}
\end{gather}
for arbitrary $\sigma$ and $\mu$.\ To accomplish this, we will make repeated use of Remark 3:
\begin{gather}
\textrm{the l.h.s. of }(\ref{eq:coefficientconjugation})=-\frac{\displaystyle
k_\sigma L_{\sigma'}\prod_{r=1}^{n-1}c_{n+\sigma_r}^\dagger 
}
{\displaystyle
L_\sigma k_{\sigma'}\prod_{r=1}^{n-1}c_{\sigma_r}^{-1} 
}=-\frac{k_\sigma L_{\sigma'}}{L_\sigma k_{\sigma'}}\prod_{r=1}^{n-1}c_{n+\sigma_r}^\dagger c_{\sigma_r} \label{eq:conj3}
\end{gather}
Note that $c_j$ can be rewritten as,
\begin{gather}
c_j=\left\{
\begin{array}{ll}
\ds\frac{L_{\{j\}\to \{n+1\cdots,2n\}}}{k_{\{j\}\to \{n+1\cdots,2n\}}}=\frac{L_{\{j\}\to \{n+1,\cdots,2n\}\backslash\{j\}}}{k_{\{j\}\to \{n+1,\cdots,2n\}\backslash\{j\}}}&(j=1,2,\cdots,n)\\
&\\
\ds -\frac{L_{\{j\}\to \{n+1,\cdots,2n\}\backslash\{j\}}}{k_{\{j\}\to \{n+1,\cdots,2n\}\backslash\{j\}}}&(j=n+1,n+2,\cdots,2n)
\end{array}\right.
\end{gather}
Thus, we have
\begin{gather}
\begin{array}{l}
\ds \prod_{r=1}^{n-1}c_{n+\sigma_r}^\dagger c_{\sigma_r}\\
\ds \ \ =\prod_{r=1}^{n-1}\left\{-\left(\frac{L_{\{n+\sigma_r\}\to \{n+1,\cdots,2n\}\backslash\{n+\sigma_r\}}}{k_{\{n+\sigma_r\}\to \{n+1,\cdots,2n\}\backslash\{n+\sigma_r\}}}\right)^\dagger\frac{L_{\{\sigma_r\}\to \{n+1,\cdots,2n\}\backslash\{\sigma_r\}}}{k_{\{\sigma_r\}\to \{n+1,\cdots,2n\}\backslash\{\sigma_r\}}}\right\}\\
\ds \ =(-)^{n-1}\prod_{r=1}^{n-1}\frac{L_{\{\sigma_r\}\to \{1,\cdots,n\}\backslash\{\sigma_r\}}}{k_{\{\sigma_r\}\to \{1,\cdots,n\}\backslash\{\sigma_r\}}}\, \frac{L_{\{\sigma_r\}\to \{n+1,\cdots,2n\}\backslash\{\sigma_r\}}}{k_{\{\sigma_r\}\to \{n+1,\cdots,2n\}\backslash\{\sigma_r\}}}\\
\ds \ =(-)^{n-1}\prod_{r=1}^{n-1}\frac{L_{\sigma_r\to \{1,\cdots,2n\}\backslash\{\sigma_r\}}}{k_{\sigma_r\to \{1,\cdots,2n\}\backslash\{\sigma_r\}}}=(-)^{n-1}\left(\frac{L_\sigma}{k_\sigma}\right)^2\frac{L_{\sigma\to \sigma'}}{k_{\sigma\to \sigma'}}
\end{array}.\label{eq:conj4}
\end{gather}
Substituting (\ref{eq:conj4}) into the r.h.s. of (\ref{eq:conj3}), we have,
\begin{gather}
\begin{array}{ll}
\displaystyle\textrm{the r.h.s. of (\ref{eq:conj3})}&\ds =-(-)^{n-1}\frac{k_\sigma L_{\sigma'}}{L_\sigma k_{\sigma'}}\left(\frac{L_\sigma}{k_\sigma}\right)^2\frac{L_{\sigma\to \sigma'}}{k_{\sigma\to \sigma'}}\\
&\ds =(-)^n\frac{L_{\{1,\cdots,2n\}}}{k_{\{1,\cdots,2n\}}}
\end{array}.
\label{eq:conj7}
\end{gather}
We also have
\begin{gather}
\begin{array}{ll}
\textrm{the r.h.s. of }(\ref{eq:coefficientconjugation})&=\frac{\displaystyle
k_\mu L_{\mu'}\prod_{r=1}^{n-1}c_{n+\mu_r}^\dagger 
}
{\displaystyle
L_\mu k_{\mu'}\prod_{r=1}^{n-1}c_{\mu_r}^{-1} 
}\ds =\frac{k_\mu L_{\mu'}}{L_\mu k_{\mu'}}\prod_{r=1}^{n-1}c_{n+\mu_r}^\dagger c_{\mu_r}\\
&\ds =\frac{k_\mu L_{\mu'}}{L_\mu k_{\mu'}}\times (-)^{n}\left(\frac{L_\mu}{k_\mu}\right)^2\frac{L_{\mu\to \mu'}}{k_{\mu\to \mu'}}=(-)^n\frac{L_{\{1,\cdots,2n\}}}{k_{\{1,\cdots,2n\}}}.
\end{array} \label{eq:conj8}
\end{gather}
Hence, from (\ref{eq:conj7}), (\ref{eq:conj8}) we see that $w_{21}^\dagger=w_{12}$.\ 
\end{proof}

\label{lastpage}

\end{document}